\documentstyle[aps,12pt,epsfig,manuscript,amsbsy]{revtex}
\global\firstfigfalse
\begin{document}
\preprint{INJE-TP-99-8, hep-th/9911193}
\def\overlay#1#2{\setbox0=\hbox{#1}\setbox1=\hbox to \wd0{\hss #2\hss}#1%
\hskip -2\wd0\copy1}

\title{Greybody factor for D3-branes in B field}

\author{ Y.S. Myung, Gungwon Kang, and H.W. Lee}
\address{Department of Physics, Inje University, Kimhae 621-749, Korea}

\maketitle

\begin{abstract}
We calculate the effect of noncommutative spacetime on the greybody 
factor on the supergravity side.
For this purpose we introduce a system of D3-branes with a 
constant NS $B$-field along their 
world volume directions ($x_2, x_3$).
Considering the propagation of minimally coupled scalar 
with non-zero momentum along($x_2, x_3$),
we derive an exact form of the greybody factor in
$B$ field.
It turns out that $\sigma^{B\ne0}_l > \sigma^{B=0}_l$.
This means that the presence of $B$-field (the noncommutativity)
suppresses the potential barrier surrounding the black hole.
As a result, it comes out the increase of greybody factor.
\end{abstract}

\newpage
Recently noncommutative geometry has attracted much interest in studying 
on string and M-theory in the 
$B$-field\cite{Con98JHEP02003,Dou98JHEP02008,She99PLB119,Has9907166,Mal9908134,Sei9908142,Big9908056}.
For simplicity, we consider supergravity solutions which 
are related to D3 branes with NS $B$ fields.
According to the AdS/CFT correspondence\cite{Mal98ATMP231}, 
the near horizon geometry 
of D=7 black hole solution can describe the large $N$ limit of 
noncommutative super Yang-Mills theory (NCSYM).
We take a decoupling limit to isolate the near horizon geometry 
from the remaining one.
It turns out that the noncommutativity affects the ultra violet(UV) but not 
the infra red(IR) of the Yang-Mills dynamics.
The NCSYM is thus not useful for studying the theory at short distances.
It is well known that an NCSYM with the noncommutativity scale $\Delta$ 
on a torus of size $\Sigma$ is equivalent to an ordinary supersymmetric 
Yang-Mills theory (OSYM) with a magnetic flux if 
$\Theta = \Delta^2/\Sigma^2$ is a rational number\cite{Has9911057}.
The equivalence between the NCSYM and the OSYM can be understood
from the T-duality of the corresponding string theory.
Hence the OSYM with $B$-field is the proper description in the UV region, 
while the NCSYM takes over in the IR region. 
Actually, the noncommutativity comes from the $B\to\infty$ limit 
of the ordinary theories\cite{Mal9908134,Sei9908142,Myu9911031}.

On the other hand, it turns out that the total number of physical 
degrees of freedom for the noncommutative case at any given 
scale coincides with the commutative case\cite{Mal9908134}.
All thermodynamic quantities of the NCSYM including the entropy are the 
same as those of the OSYM.
However, in the next order correction of the $\alpha'$-expansion, 
the entropy decreases in the NCSYM\cite{Cai9910092}.
We remind the reader that aside the entropy, there exists an 
important dynamical quantity ``the greybody factor(absorption cross section)'' 
for the quantum black 
hole\cite{Gub97NPB217,Gub99CMP325,Gub98NPB393,Lee98PRD104013}.
Hence it is very important to check whether there is or not a change 
in the greybody factors between the commutative and the 
noncommutative cases.
On the string side, there was a calculation for the absorption of 
scalars into the noncommutative D3-branes\cite{Hyu9909059}.

In this paper we wish to study the quantum aspects of D3-brane black hole
in $B$-field background using a minimally coupled scalar.
This may belong to the off-diagonal gravitons polarized parallel to the brane 
($h_{ab}, a,b=0,1,2,3$).
We will derive an exact form of absorption cross section 
in $B$-field on the supergravity side. 
However, it is not easy to obtain the absorption cross section 
on the gauge theory side because the graviton can turn into pairs of 
gauge bosons, scalars, or fermions in the world volume approach.
In the absence of $B$-field, adding up these rates leads to 
the absorption cross section of $h_{ab}$ on the superstring 
side\cite{Gub97NPB217}.

We start with the D=7 extremal black hole solution for D3-branes 
in a $B_{23}$-field on the D=10 string frame\cite{Mal9908134}
\begin{eqnarray}
ds_{str}^2 &=& f^{- {1 \over 2}} \left \{ -dx_0^2 + dx_1^2 
      + h ( dx_2^2 + dx_3^2 ) \right \} 
     + f^{ {1 \over 2}} \left ( dr^2 + r^2 d\Omega_5^2 \right ),
   \label{metric-str}  \\
f &=& 1 + {{R_B^4} \over r^4 }, 
     ~~ h^{-1} = \sin^2 \theta f^{-1} + \cos^2 \theta, 
     \nonumber \\
B_{23} &=& \tan \theta f^{-1} h, ~~ e^{2 \phi} = g^2 h, \nonumber \\
F_{01r} &=& {1 \over g} \sin \theta \partial_r(f^{-1}),
    ~~ F_{0123r} = {1 \over g} \cos \theta h \partial_r(f^{-1}).
   \nonumber
\end{eqnarray}
Here the asymptotic value of $B$-field is $B_{23}^\infty = \tan \theta$ and
the parameter $R$ is defined by $\cos \theta R_B^4 = 4 \pi g N \alpha'^2$ 
with $N$(the number of D3-branes). 
Here we define $R_0^4 = 4 \pi g N \alpha'^2$ for $B=0$, 
$R_\infty$ for $B\to\infty$ and $R_B$ for an arbitrary $B$.
We note that $R_B > R_0$.
And $g=g_\infty$ is the string coupling constant.
It is obvious that for $\theta=0$, one recovers the ordinary D3-brane 
black hole with the standard AdS$_5\times$S$^5$ in the near horizon. 
In this case the dilaton background($\phi$) is constant.
But for $\theta=\pi/2$ one finds the D3-brane black hole in the 
$B\to\infty$ limit.
This actually reduces to the black D-string with the smeared 
coordinates $x_2$ and $x_3$.
In latter case, one finds a deviation from AdS$_5\times$S$^5$ in the 
near horizon.

Now let us introduce the perturbation analysis to derive the greybody factor.
General fluctuations including the dilaton and the RR  
scalar will be given as the complicated 
systems of differential equations\cite{Myu9912288}.
A simple equation may be arised from the off-diagonal 
graviton fluctuation of $h_{01}$\cite{Mal98ATMP231,Gub97NPB217}.
Let us set all other fluctuations to zero and introduce a 
minimally coupled scalar $\varphi$ which may be $g^{00} h_{01}$.
This scalar is a good test field in the Einstein frame
because it is minimally coupled to the background spacetime.
Then this in the Einstein frame satisfies
\begin{equation}
{{ e^{2 \phi} } \over \sqrt{-g} } \partial_M \left ( \sqrt{-g} 
     e^{-2 \phi} g^{MN} \partial_N \varphi \right ) =0.
\label{equationofmotion}
\end{equation}
This leads to
\begin{equation}
{1 \over \sqrt{-g} } \partial_M \left ( \sqrt{-g} g^{MN} 
      \partial_N \varphi \right ) -
     2 \left ( \partial_M \phi \right ) 
       \left ( \partial^M \varphi \right ) = 0,
\label{reducedequation}
\end{equation}
where $g_{MN}$ is the string frame metric in (\ref{metric-str}).

Now let us consider 
\begin{equation}
\varphi(t,x_1, x_2, x_3, r, \theta_i ) = e^{-i \omega t} 
     e^{i(k_1 x_1 + k_2 x_2 + k_3 x_3 )}
     Y_l(\theta_1, \theta_2, \cdots, \theta_5) \varphi^l(r) 
\label{defphi}
\end{equation}
with $\bar \nabla^2_{\theta_i} Y_l(\theta_i) = - l(l+4)Y_l(\theta_i)$.
Here $\varphi^l(r)$ is the radial part of the $l$-partial wave of 
energy $\omega$.
We assume the momentum dependence along the world volume directions.
This is a crucial step to translate the noncommutativity in the 
$x_2 , x_3$ directions(the presence of $B_{23}$) 
into the greybody factor\cite{Mal9908134}.
Then Eq. (\ref{reducedequation}) takes the form
\begin{equation}
\left \{ 
{ \partial^2 \over {\partial r^2}} + { 5 \over r} {\partial \over {\partial r}}
  - {{l(l+4)} \over r^2 } + (\omega^2 - k_1^2 ) f 
  - (k_2^2 + k_3^2 ) \left ( 
  \cos^2\theta(f-1) + 1 \right )
\right \} \varphi^l =0 .
\label{lequation}
\end{equation}
If $\theta=0$ ($B$-field is turned off), one finds the ordinary 
scalar propagation in the D=7 black hole spacetime\cite{Gub99CMP325}
\begin{equation}
\left \{ 
{ \partial^2 \over {\partial r^2}} + { 5 \over r} {\partial \over {\partial r}}
  - {{l(l+4)} \over r^2 } + (\omega^2 - k_1^2  
  - k_2^2 - k_3^2 ) \left ( 1 + {R_0^4 \over r^4} \right )  
\right \} \varphi^l_{B =0} =0 .
\label{lB0equation}
\end{equation}
On the other hand, in the presence of the 
$B\to\infty$ limit ($\theta \to \pi/2$), Eq.(\ref{lequation}) leads to
\begin{equation}
\left \{ 
{ \partial^2 \over {\partial r^2}} + { 5 \over r} {\partial \over {\partial r}}
  - {{l(l+4)} \over r^2 } + (\omega^2 - k_1^2 ) f 
  - (k_2^2 + k_3^2 ) 
\right \} \varphi^l_{B \to \infty} =0 .
\label{lBequation}
\end{equation}
For simplicity, we require $k_1=0$, $k_2, k_3 \ne0$. 
And  the low energy limit of 
$\omega \to 0$ but still $\omega^2 > k_2^2 + k_3^2$.
Fortunately, the absorption cross section for the 
$l$-partial wave for Eq. (\ref{lB0equation}) can be extracted from the 
solution to Mathieu's equation\cite{Gub99CMP325},
\begin{equation}
\sigma_l^{B=0} = {{ 8 \pi^2/3} \over \omega^5} (l+1)(l+2)^2 (l+3) P_l,
\label{sigmaB0}
\end{equation}
where the absorption probability $P_l$ takes the form
\begin{equation}
P_l = {{4 \pi^2 } \over { \left [ (l+1)! \right ]^4 (l+2)^2 } }
   \left ( {{\tilde \omega R_0} \over 2 } \right )^{8+4l} 
  \sum_{n=0}^\infty \sum_{k=0}^n b_{n,k}^l (\tilde \omega R_0)^{4n} 
     (\log \tilde \omega \bar R_0 )^k.
\label{absprob}
\end{equation}
Here $b_{0,0}^0=1$, $b_{1,1}^0 = -{1 \over 6}$, $b_{1,0}^0 = {7 \over 72}$,
$\bar R_0 = e^\gamma R_0/2$ with $\gamma=0.5772$ (Euler's constant), and 
$\tilde \omega = \sqrt{\omega^2 - k_2^2 - k_3^2} $ 
$\simeq \omega ( 1 - {{k_2^2 + k_3^2} \over {2 \omega^2}} )$.
In the case of $l=0$ mode ($s$-wave), one finds
\begin{equation}
\sigma_0^{B=0} = {{\pi^4 (\tilde \omega R_0)^8 } \over { 8 \omega^5 } }
  \left \{
  1 - { 1\over 6} \left ( \tilde \omega R_0 \right )^4 
       \log (\tilde \omega \bar R_0 ) + 
      { 7 \over 72} (\tilde \omega R_0)^4 + \cdots 
  \right \}
\label{sigmaB0l0}
\end{equation}
in the low-energy approximation of $\tilde \omega R_0 < 1$.
In this case we see that the logarithmic term is greater than 
the fourth power order term.

On the other hand, for $l=0$, $\rho=\hat \omega r$, Eq.(\ref{lBequation})
leads to
\begin{equation}
\left \{
{{\partial^2} \over {\partial \rho^2}} 
+ { 5 \over \rho} { \partial \over {\partial \rho}}  + 1
+ {{(\hat \omega R_\infty)^4} \over \rho^4 } 
\right \} \varphi^0_{B \to \infty} =0
\label{rhoBequation}
\end{equation}
with $\hat \omega^2 = \tilde \omega \omega $.
Let us introduce the new coordinate $y$ and function $\psi$ as 
\begin{equation}
y = {{(\hat \omega R_\infty)^4} \over \rho } ~~{\rm and} ~~
\varphi^0_{B \to \infty}(\rho) = y^4 \psi_{B \to \infty}(\rho).
\label{defpsi}
\end{equation}
Then one finds a dual equation to Eq.(\ref{rhoBequation}) which is 
written in terms of $y$ as
\begin{equation}
\left \{
{{\partial^2} \over {\partial y^2}}
+ { 5 \over y} {\partial \over {\partial y}} + 1
+ {{(\hat \omega R_\infty)^4} \over y^4}
\right \} \psi_{B \to\infty}(\rho) = 0.
\label{dualequation}
\end{equation}
Using a trick for an improved matching of inner and outer solutions 
in Ref. \cite{Gub98NPB393},
one finds the absorption cross section
\begin{equation}
\sigma^{B\to \infty}_{0}={{\pi^4 (\hat \omega R_\infty)^8} \over { 8 \omega^5}} 
\left \{ 1 - {1 \over 6} (\hat \omega R_\infty)^4 \log(\hat \omega R_\infty ) + 
{\cal O}((\hat \omega R_\infty)^4)
\right \}
\label{sigmaBl0}
\end{equation}
with $\hat \omega = \sqrt{\omega \tilde \omega}$
$= \omega \left ( 1 - {{k_2^2 + k_3^2} \over \omega^2 } \right )^{1/4}$
$\simeq \omega \left ( 1 - {{k_2^2 + k_3^2 } \over {4 \omega^2}} \right )$.

Although $\sigma_0^{B=0}$ and $\sigma_0^{B\to\infty}$ take the same
form up to the leading-correction, there exist some differences to point out.
First, because of $\tilde \omega < \hat \omega$ and 
$R_0 < R_\infty$, one finds
$\sigma_0^{B=0}<\sigma_0^{B\to\infty}$.
This implies that in the presence of large $B$-field($B\to\infty$),
the height of potential is lower than the case of $B=0$.
Actually, we observe from Eq.(\ref{lequation}) that
the height of the effective potential decreases as
$\theta$ (that is, the strength of $B$-field) increases.
Hence, in the low energy scattering, one finds the larger
absorption cross section $\sigma_0^{B\to\infty}$ than $\sigma_0^{B=0}$.
The greybody factor of the black hole arises as a consequence of
scattering of minimally coupled scalar off the gravitational
potential barrier surrounding the horizon; that is, this is an effect
of spacetime curvature\cite{Lee98PRD104013}.
Furthermore, we wish to point out that there also exists the difference 
for ``$R$'' in the logarithmic terms.

We have seen the effect of $B\to\infty$ on the absorption probability 
for D3-branes above.
Now we may ask how the greybody factor changes in the presence of arbitrary 
$B$-field.
For an arbitrary $B$, we can rewrite the Eq.(\ref{lequation})  
in the following form: 
\begin{equation}
\left \{ 
{ \partial^2 \over {\partial r^2}} + { 5 \over r} {\partial \over {\partial r}}
  - {{l(l+4)} \over r^2 } + 
  \tilde \omega^2 \left ( 1 +  
  { {\tilde R_B^4} \over r^4} \right )
\right \} \varphi^l_B =0, 
\label{Bequation}
\end{equation}
where $\tilde R_B^4 = \left \{ 1 + {{(k_2^2 + k_3^2 ) \sin^2 \theta } 
\over {\tilde \omega^2}} \right \} R_B^4 $.
Suprisingly, we find that the above equation has exactly the same form 
as Eq.(\ref{lB0equation}) with different ``$R$''.
Then the absorption cross section can be read off from 
(\ref{sigmaB0}) by substituting $R_0$ with $\tilde R_B$ as follows
\begin{equation}
\sigma_l^B = \sigma_l^{B=0}(R_0 \to \tilde R_B ).
\label{sigmaB}
\end{equation}
This is our main result.
From (\ref{sigmaB}) one can recover $\sigma_0^{B=0}$ in (\ref{sigmaB0l0}) 
for $\theta=0$ as well as 
$\sigma_0^{B\to\infty}$ in (\ref{sigmaBl0}) for $\theta=\pi/2$.
Here we point out several main features.
Firstly, for arbitrary $B$-field, 
one finds $\sigma_l^{B\ne 0} > \sigma_l^{B=0}$ since 
$\tilde R_B > R_0$ and $\tilde \omega \tilde R_B < 1$.
Secondly, since $\tilde R_B$ incerases as $\tilde \omega \to 0$, 
the wave packet having lower energy is absorbed more easily.  
Thirdly, the absorption cross section in the presence of 
$B$-field is exactly same as that in the absence of $B$-field
with $\tilde R_B = \left \{ 1 + {{(k_2^2 + k_3^2 ) \sin^2 \theta }
\over {\tilde \omega^2}} \right \}^{1/4} R_B $. 
It may imply that the scattering wave regards 
the geometry as being the same with an increased AdS$_5$ radius as long as 
the absorption cross section is concerned.
Furthermore, a recent paper\cite{Das9911137} pointed out that the 
Einstein metric of a D3-brane is not changed by the presence of a 
$B$-field.
This implies that for a minimally coupled scalar, the solution with 
the $B$-field can be obtained from the solution without the 
$B$-field by replacing $R_0$ by $\tilde R_B$.

We conclude that the main effect of $B$-field (the noncommutativity) 
suppresses the potential barrier surrounding the black hole.
As a result, one finds the increase of greybody factor.
Finally we comment on the relationship between the near horizon geometry and 
the greybody factor.
Taking the decoupling limit\cite{Mal9908134}, the near-horizon
geometry(a deviation from AdS$_3\times$S$^5$ for $B\ne0$) plays 
an important role in obtaining two-point function of 
$\varphi=g^{00} h_{01}$ : 
$\langle T_{01}(k) T_{01}(-k) \rangle \simeq \exp(-c a |k|)$
with $c=1.69, a = \sqrt{\tilde b R_B^2}$.
In the UV regime, this decays exponentially with the momentum 
$k=\sqrt{k_2^2 + k_3^2}$.
This is a new feature in the noncommutative geometry.
However we remind the reader that in the calculation of the greybody factor
we never take the decoupling limit to single the near horizon 
geometry out.
Instead we use the differential equation to find 
the approximate solution in the whole region.
Hence the near horizon geometry is not so sensitive to 
obtain the greybody factor.
In our case, the momentum contributes to the cross section through 
$\tilde R_B^4 = \{ 1 + {k^2\over \tilde \omega^2} \sin^2\theta \} R_B^4$
with a sequence $\tilde R_B^4 > R_B^4 > R_0^4$.

\vspace*{2.0cm}
\noindent
{\it Note Added}\\
After our work has been done, we find a related 
paper \cite{Kay9911183}. The first version of this paper takes over the same 
subject with the RR scalar ($C$), which is non-minimally coupled 
to the background. 
Kaya claims that the greybody factor does not 
change even if the large $B$-field is turned on. 
However, we point out that such result 
comes from the assumption that the scalar field $C$ does not depend on 
the noncommuting coordinates ($x_2$, $x_3$). 
In our case, if $\varphi$ does not depend on ($x_2$, $x_3$), 
equivalently, $k_2 = k_3 =0$, we also recover the usual result for 
ordinary D3-branes except with $R_B > R_0$. 
It indicates that, in order to extract 
the B-field effect(the noncommutativity), 
the world volume dependence of the scalar field is a key 
ingredient\cite{Mal9908134,Das9911137}. 
It follows because the propagation of fields does not 
feel the presence of B-field if the wave does not move 
on the $x_2$-$x_3$ plane\cite{Myu9911031}. 
Fortunately, Kaya in the third version of ref.\cite{Kay9911183} 
comment that $\sigma^{B \ne 0}_{0} > \sigma^{B=0}_0$ is still 
valid for the case of the RR scalar by introducing the new
parameter $s =\sqrt{ 1 - k^2/\omega^2} < 1$.
In our notation, 
$\tilde \omega = s \omega$ and $\hat \omega = \sqrt{s} \omega$.
That is, he showed that 
$\sigma_0^{B\ne 0} \propto s^4 > \sigma_0^{B=0} \propto s^8$. 

\section*{Acknowledgement}
We are greteful to Sumit R. Das for helpful comments.
This work was supported by the Brain Korea 21
Program, Ministry of Education, Project No. D-0025.


\begin{references}
\bibitem{Con98JHEP02003}
  A. Connes and M. Douglas, JHEP {\bf 9802}, 003(1998), hep-th/9711162.
\bibitem{Dou98JHEP02008}
  M. Douglas and C. Hull, JHEP {\bf 9802}, 008(1998), hep-th/9711165.
\bibitem{She99PLB119}
M.M. Sheikh-Jabbari, Phys. Lett. {\bf B450}, 119(1999), hep-th/9810179.
\bibitem{Has9907166}
  A. Hashimoto and N. Itzhaki, hep-th/9907166.
\bibitem{Mal9908134}
  J. Maldacena and J. Russo, hep-th/9908134.
\bibitem{Sei9908142}
  N. Seiberg and E. Witten, hep-th/9908142.
\bibitem{Big9908056}
  D. Bigatti and L. Susskind, hep-th/9908056.
\bibitem{Mal98ATMP231}
  J. Maldacena, Adv. Theor. Math. Phys. {\bf 2}, 231(1998), hep-th/9711200.
\bibitem{Has9911057}
  A. Hashimoto and N. Itzhaki, hep-th/9911057.
\bibitem{Myu9911031}
  Y.S. Myung and H.W. Lee, hep-th/9911031; hep-th/9910083.
\bibitem{Cai9910092}
  R.G. Cai and N. Ohta, hep-th/9910092.
\bibitem{Gub97NPB217}
  S. Gubser, I. Klebanov, and A. Tseytlin, Nucl. Phys. {\bf B499}, 217(1997),
      hep-th/9703040;
  S. Gubser and I. Klebanov, Phys. Lett. {\bf B413}, 41(1997), hep-th/9708005;
  S. Gubser, I. Klebanov, and A. Polyakov, Phys. Lett. {\bf B428}, 105(1998),
       hep-th/9802109.
\bibitem{Gub99CMP325}
  S. Gubser and A. Hashimoto, Comm. Math. Phys. {\bf 203}, 325(1999); 
        hep-th/9805140.
\bibitem{Gub98NPB393}
  S. Gubser, A. Hashimoto, I. Klebanov, and M. Krasnitz, 
      Nucl. Phys. {\bf B526}, 393(1998), hep-th/9803023.
\bibitem{Lee98PRD104013}
  H.W. Lee and Y.S. Myung, Phys. Rev. {\bf D58}, 104013(1998), hep-th/9804095;
  hep-th/9903054; H.W. Lee, N.J. Kim, and Y.S. Myung, hep-th/9805050.
\bibitem{Hyu9909059}
  S. Hyun, Y. Kiem, S. Lee, and C.Y. Lee, hep-th/9909059.
\bibitem{Myu9912288}
  Y.S. Myung, G. Kang, and H.W. Lee, hep-th/9912288.
\bibitem{Das9911137}
  S.R. Das, S.K. Rama, and S.P. Trivedi, hep-th/9911137.
\bibitem{Kay9911183}
  A. Kaya, hep-th/9911183.
\end{references}
\end{document}